# Compensation of nonlinear phase shifts with third-order dispersion: fiber stretchers can out-perform grating stretchers in short-pulse fiber amplifiers

**Shian Zhou, Lyuba Kuznetsova, Andy Chong, and Frank W. Wise**

*Department of Applied Physics, Cornell University, Ithaca, NY 14853*
*Sz53@cornell.edu*

**Abstract:** We show that nonlinear phase shifts and third-order dispersion can compensate each other in short-pulse fiber amplifiers. In particular, we consider chirped-pulse fiber amplifiers at wavelengths for which the fiber dispersion is normal. The nonlinear phase shift accumulated in the amplifier can be compensated by the third-order dispersion of the combination of a fiber stretcher and grating compressor. A numerical model is used to predict the compensation, and initial experimental results that exhibit the main features of the calculations are presented. In the presence of third-order dispersion, an optimal nonlinear phase shift reduces the pulse duration, and enhances the peak power and pulse contrast compared to the pulse produced in linear propagation. Contrary to common belief, fiber stretchers can perform as well or better than grating stretchers in fiber amplifiers, while offering the major practical advantages of a waveguide medium. The relative benefits of a fiber stretcher increase with increasing pulse energy, so the results presented here will be relevant to fiber amplifiers designed for the highest pulse energies.

@2005 Optics Society of America

**OCIS codes:** (190.7110) Ultrafast nonlinear optics; (320.7140) Ultrafast processes in fibers; (260.2030) Dispersion

___

**References and links**
1. G. P. Agrawal, *Nonlinear Fiber Optics* 3$^{rd}$ (Academic, San Diego, California, 2001).
2. D. Strickland and G. Mourou, "Compression of amplified chirped optical pulses," *Opt. Commun*. **56**, 219-221 (1985).
3. M. Pessot, P. Maine and G. Mourou, "1000 times expansion/compression of optical pulses for chirped pulse amplification," *Opt. Commun*. **62**, 419-421 (1987).
4. O. E. Martinez, "Design of high-power ultrashort pulse amplifiers by expansion and recompression," *IEEE J. Quantum. Electron*. **23**, 1385-1387 (1987).
5. M. D. Perry, T. Ditmire, and B. C. Stuart, "Self-phase modulation in chirped pulse amplifiers," *Opt. Lett.* **19**, 2149-2151 (1994).
6. A. Braun, S. Kane and T. Norris, "Compensation of self-phase modulation in chirped-pulse amplification laser system," *Opt. Lett.* **22**, 615-617 (1997)
7. O. E. Martinez, "3000 times grating compressor with positive group velocity dispersion: application to fiber compensation in the 1.3-1.6 μm region," *IEEE J. Quantum Electron*. **23**, 59-64 (1987).
8. S. Kane and J. Squier, "Grating compensation of third-order material dispersion in the normal-dispersion regime: sub-100-fs chirped-pulse amplificiation using a fiber stretcher and grating-pair compressor," *IEEE J. Quantum Electron*. **31**, 2052-2058 (1995).
9. For a recent example, see G. Imeshev, I. Hartl, and M. E. Fermann, "Chirped pulse amplification with a nonlinearly chirped fiber Bragg grating matched to the Treacy compressor," Opt. Lett. 29, 679-681, (2004). See also references therein.
10. J. Limpert, A. Liem, T. Schreiber, M. Reich, H. Zellmer, A. Tunnerman, "High-performance ultrafast fiber laser systems," in Fiber Lasers: Technology, Systems, and Applications, ed. L. N. Durvasula, *Proceedings of SPIE* vol. 5335 (SPIE, Bellingham, WA, 2004), pp. 245-252.
11. M. E. Fermann, A.Galvanauskas, and M. Hofer, "Ultrafast pulse sources based on multi-mode optical fibers," Appl. Phys. B **70**, 1, (2000).
12. H. Lim, F. Ö. Ilday, and F. W. Wise "Generation of 2 nJ pulses from a femtosecond Yb fiber laser," Opt. Lett. **28**, 660-662 (2003).
13. H. Lim, J. Buckley, and F. W. Wise, "Wavelength tunability of femtosecond Yb fiber lasers," Conference on Lasers and Electro-Optics (Optical Society of America, San Francisco, Calif., 2004), presentation CThK3.
14. R. Trebino and D. J. Kane, "Using phase retrieval to measure the intensity and phase of ultrashort pulses: frequency-resolved optical gating", *J. Opt. Soc. Am. A*, **10**, 1101-1111 (1993).

___

## 1. Introduction

It is well-known that nonlinear phase shifts ($\Phi^{NL}$) can lead to distortion of short optical pulses [1]. In chirped-pulse amplification (CPA) [2], a pulse is stretched to reduce the detrimental nonlinear effects that can occur in the gain medium. After amplification, the pulse is dechirped, ideally to the duration of the initial pulse. The stretching is typically accomplished by dispersively broadening the pulse in a segment of fiber or with a diffraction-grating pair. For pulse energies of microjoules or greater, the dechirping is done with gratings, to avoid nonlinear effects in the presence of anomalous group-velocity dispersion (GVD), which are particularly limiting. The magnitude of the dispersion of a grating stretcher can exactly equal that of the gratings used to dechirp the pulse, to all orders [3, 4]. At low energy, the process of stretching and compression can thus be perfect. At higher energy, some nonlinear phase will be accumulated and this will degrade the temporal fidelity of the amplified pulse. For many applications, $\Phi^{NL}$ (also referred to as the B-integral) must be less than 1 to avoid unacceptable structure on the amplified pulse [5, 6].

The total dispersion of a fiber stretcher differs from that of a grating pair, and this mismatch results in uncompensated third-order dispersion (TOD), which will distort and broaden the pulse, at least in linear propagation. At wavelengths where the fiber has normal GVD (such as 1 μm, which will be our main focus here), the TOD of the fiber adds to that of the grating pair. Stretching ratios of thousands are used in CPA systems designed to generate microjoule and millijoule-energy pulses, in which case the effects of TOD would limit the dechirped pulse duration to the picosecond range. It has thus become "conventional wisdom" that fiber stretchers are unacceptable in CPA systems, and as a consequence, grating stretchers have become ubiquitous in these devices.

Apart from the difficulty of compensating the cubic phase, a fiber offers major advantages as the pulse stretcher in a CPA system. Coupling light into a fiber is trivial compared to aligning a grating stretcher. The grating stretcher includes an imaging system [7] that can be misaligned, and when misaligned will produce spatial and temporal aberrations in the stretching. A fiber stretcher cannot be misaligned. The fiber is also less sensitive to drift or fluctuations in wavelength or the pointing of the beam that enters the stretcher. Beam-pointing fluctuations may reduce the coupling into the fiber to below the optimal level, whereas they translate into changes in dispersion with a grating pair. Finally, the spatial properties of the beam can influence the stretching with a grating pair, while the pulse stretched in a fiber cannot have spatial chirp, experiences the same stretching at all points on the beam, and exits the fiber with a perfect transverse mode. With fiber amplifiers, there is naturally strong motivation to employ fiber stretchers – grating stretchers detract substantially from the benefits of fiber.

One possible solution to this problem is the combination of a fiber stretcher with a grism pair for dechirping, as proposed by Kane and Squier [8]. However, grisms require a challenging synthesis and to date have not found significant use. Recently, significant attention has been devoted to the development of fiber Bragg gratings (FBGs), including chirped FBGs, for use in CPA systems [9]. A chirped fiber grating designed to compensate higher-order dispersion is conceptually the same as a diffraction-grating stretcher, which is matched to the compressor to all orders, but it offers the practical advantages of fiber discussed above. There is no report in the literature of a fiber grating designed to compensate nonlinearity. The authors of Reference 9 point out that the output pulse is apparently distorted owing to the nonlinear phase shift at the highest energies. It is an experimental fact that the fiber CPA systems that produce the highest pulse energies to date [10, 11] employ ordinary diffraction gratings, not chirped fiber gratings.

Here we show that the nonlinear phase shift accumulated by a pulse in amplification can actually be compensated to some degree by the cubic spectral phase. (Or *vice-versa* -- the cubic spectral phase can be compensated by nonlinearity.) This conclusion has important consequences for short-pulse fiber amplifiers. First, *the performance of a CPA system with a fiber stretcher and grating compressor actually improves with nonlinearity*. This behavior contrasts with that of a CPA system with grating stretcher and compressor, where the pulse fidelity decreases monotonically with nonlinear phase shift. Second, near the optimal value of the nonlinear phase, *a fiber stretcher can perform better than a grating stretcher*. This compensation of TOD by nonlinear phase shift will not offer much advantage in solid-state amplifiers, in which the spatial consequences of the nonlinear phase shift will be limiting: small-scale or whole-beam self-focusing will distort the beam or, eventually, damage the gain medium. Most of the effort to develop CPA systems has involved bulk solid-state gain media, as opposed to fibers. This may explain why the process described here has not been pointed out previously. In a single-mode fiber, there are no transverse variations, so the longitudinal effects of the nonlinear phase can be isolated and exploited. In this report we focus on ordinary single-mode fiber, but the approach can be used in any implementation of CPA, such as those employing fiber Bragg gratings or photonic bandgap fibers.

## 2. Principle and numerical modeling

It is perhaps remarkable that nonlinear phase shifts and TOD should compensate each other. After all, TOD acting alone produces an anti-symmetric phase, while self-phase modulation alone produces a symmetric phase. We can offer a qualitative rationale for the compensation by considering pulse propagation in a fiber stretcher, followed by a nonlinear segment with negligible dispersion, and a grating compressor. After stretching, the pulse develops a slightly asymmetric intensity profile owing to the TOD. However, because GVD dominates the stretching, the mean time in the pulse remains close to zero. That is, the pulse energy does not shift in time. This implies that the peak of the pulse moves to slightly earlier (negative) time, while the asymmetric tail associated with TOD extends to later time, for $d^3\phi/d\omega^3 > 0$. The nonlinear phase shift accumulated by the stretched pulse is then also peaked at slightly negative time, and the temporal phase has negative slope at the center of the pulse. In the compressor, the large quadratic phase from GVD is subtracted off, while the positive TOD counters the negative phase slope. Analogous arguments can be made in the frequency domain, because time is approximately mapped to frequency in highly-chirped pulses. This intuitive explanation is consistent with the numerical results presented below, but more work is needed to understand this process thoroughly.

Numerical simulations were employed to study CPA with a fiber stretcher, a fiber amplifier and a grating compressor (the key elements of the experimental setup in Fig. 3). The parameters of the simulations were taken as those of the experiments described below, to allow comparison of theory and experiment. All the fiber is single-mode fiber (SMF). The input pulses to this system represent the output of an Yb fiber oscillator [12, 13] and were taken to be 150-fs gaussian pulses with 10.4-nm bandwidth at 1060 nm. The stretcher consists of 100 m of SMF, and analogous results for a 400-m stretcher will be summarized below. The amplifier consists of 1 m of Yb-doped gain fiber, and is followed by 3 m of SMF, where most of the nonlinear phase shift is accumulated. The magnitude of the nonlinear phase shift is adjusted by varying the gain of the amplifier. The compressor is a pair of gratings with 1200 lines/mm, used in a double-pass configuration. The GVD and TOD of the fibers and grating pairs are included in the simulations (the numerical values are listed in the caption of Fig. 1). The nonlinear Schrödinger equations that govern propagation in each section are solved by the standard split-step technique.

After propagation through 100 m of stretcher fiber, the pulse duration is 46 ps. The compressor grating separation is optimized to produce the shortest output pulse at each pulse energy; as expected, the grating separation decreases with increasing $\Phi^{NL}$. All numerical and experimental results reported here are based on the optimal grating separations.

As an approximation to linear propagation, the amplifier gain was adjusted to produce a low-energy (4 nJ) pulse. The resulting $\Phi^{NL} = 0.4\pi$. The pulse shape after compression (Fig. 1(a)) exhibits the signature asymmetric broadening and secondary structure from TOD. The full-width at half-maximum (FWHM) pulse duration has increased to 290 fs, and the peak power is 9.5 kW. The envelopes of the interferometric autocorrelation of the output pulse are shown in Fig. 1(b). The autocorrelations are provided because they will be compared to experimental results below.

Increasing the nonlinear phase shift improves the quality of the output pulse. Best results are obtained with amplified pulse energy of 19 nJ, which produces $\Phi^{NL} \approx 1.9\pi$. With this nonlinear phase shift, the compressed pulse (Fig. 1(c)) duration is reduced to 191 fs, which is within ~25% of the original pulse width. Equally significant is the suppression of the trailing "wing" of the pulse. The resulting peak power is 74 kW; the pulse energy is 5 times larger than in Fig. 1(a), but the peak power is 8 times larger owing to the improved pulse quality. The corresponding autocorrelation is shown in Fig. 1(d). The power spectrum broadens by less than 5% at the highest energy. Spectral broadening is roughly proportional to the nonlinear phase shift divided by the stretching ratio, so small broadening is expected for $\Phi^{NL} \sim 2\pi$ and a stretching ratio of 300. For larger nonlinear phase shifts, the pulse quality degrades. Thus, for a given amount of cubic phase, there is an optimal value of $\Phi^{NL}$. Simulations with the signs of the TOD of the fiber and grating pair reversed produce identical results; a positive (self-focusing) nonlinearity can compensate the effects of either sign of TOD.

For comparison, we show the best results that can be obtained with a grating stretcher and compressor at the same value of $\Phi^{NL}$ (*i.e.*, the same pulse energy). The compressed pulse (Fig. 1(e)) has FWHM duration 214 fs and peak power 67 kW. Thus, for this low stretching ratio and pulse energy, the fiber stretcher offers ~10% improvement over the grating stretcher. However, we emphasize that the advantage of the fiber stretcher increases with increased stretching ratio and nonlinear phase shift.

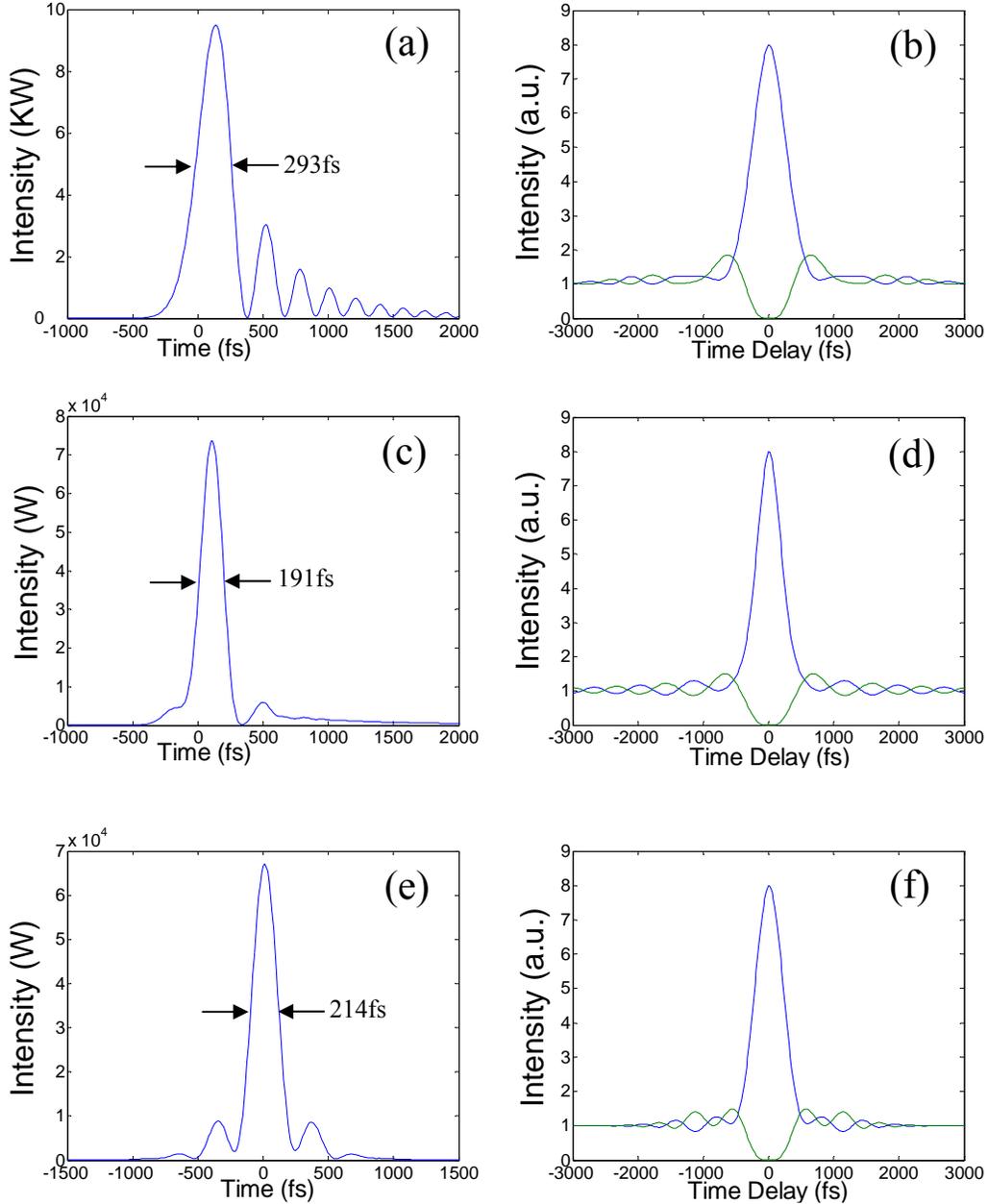

Figure 1. Intensity profiles (a, c, e) and autocorrelation envelopes (b, d, f) obtained from numerical simulations. a) and b): fiber stretcher, $\Phi^{NL} = 0.4\pi$. c) and d): fiber stretcher, $\Phi^{NL} = 1.9\pi$. e) and f) grating stretcher, $\Phi^{NL} = 1.9\pi$. Parameters used in the simulations: nonlinear coefficient $\gamma = 4.3$ kW$^{-1}$m$^{-1}$; GVD coefficient $\beta_2 = 230$ fs$^2$/cm and TOD coefficient $\beta_3 = 254$ fs$^3$/cm. For the grating-pair compressor, $\beta_2 = -1.2\times10^5$ fs$^2$/cm and $\beta_3 = 4.5\times10^5$ fs$^3$/cm.

The results of a series of similar calculations with varying nonlinear phase shift are summarized in Fig. 2(a). We define the relative peak power as the ratio of the peak power of the output pulse to the ideal peak power that would be obtained in the absence of TOD and nonlinear phase shift. That is, the ideal peak power is that of the output pulse energy with the input pulse intensity profile. With a grating stretcher, the relative peak power decreases monotonically with $\Phi^{NL}$. This trend is well-known and is the reason why $\Phi^{NL}$ is typically limited to 1 in CPA systems. The relative peak power with a fiber stretcher increases until $\Phi^{NL} \sim 2\pi$, and then decreases. For $\Phi^{NL} > 1.5\pi$ the fiber stretcher performs better than the grating stretcher, albeit by a small margin. The inset shows the variation of the FWHM pulse duration with $\Phi^{NL}$.

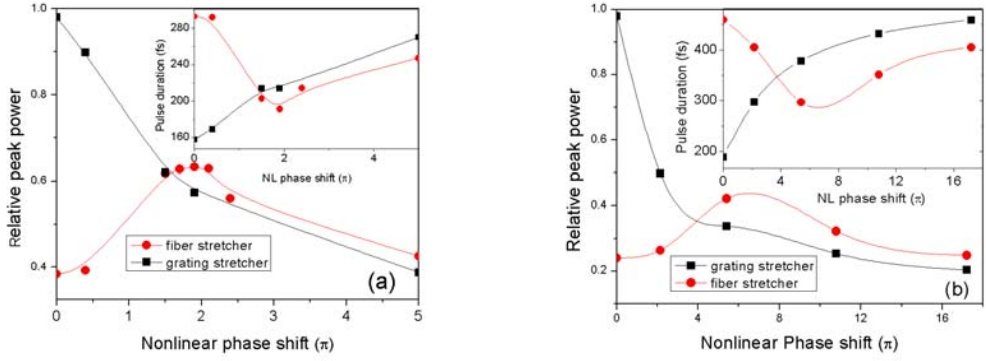

Figure 2. a) Variation of relative peak power with $\Phi^{NL}$ for CPA with grating (square symbols), and fiber (round symbols) stretchers. The pulse is stretched to ~46 ps in each case. The inset shows the variation of the pulse duration with $\Phi^{NL}$. b) Same as a) but with 400-m stretcher, which stretches to ~140 ps. The lines are only to guide the eye.

Analogous results for a 400-m stretcher are shown in Fig. 2(b) and, together with Fig. 2(a), illustrate the scaling of the compensation. The optimal value of $\Phi^{NL}$ increases roughly linearly with the magnitude of the TOD. The benefit of nonlinearity is larger with greater TOD: with a 100-m stretcher the relative peak power is ~50% larger than that obtained in linear propagation ($\Phi^{NL} = 0$), while with a 400-m stretcher the relative peak power increases by nearly a factor of two. The advantage of a fiber stretcher over a grating stretcher at the optimal value of $\Phi^{NL}$ also increases with increasing TOD. On the other hand, the maximum value of the relative peak power decreases with increasing TOD. With a 400-m stretcher, a 500-nJ pulse produces $\Phi^{NL} = 5.4\pi$. The compressed pulse duration is 300 fs (compared to ~500 fs with $\Phi^{NL} = 0$, or 380 fs with $TOD = 0$), and with 60% compressor efficiency a peak power of 1 MW is reached. If the TOD is doubled while all other parameters are held constant, the pulse energy reaches 1 µJ, with $\Phi^{NL} = 10.8\pi$. The compressed pulse duration is still 300 fs, and the peak power reaches ~2 MW.

## 3. Experimental results

A schematic of the experimental setup is shown in Fig. 3. The Yb fiber laser generates 140-fs pulses with ~12 nm bandwidth at 1060 nm. The parameters of the experiment are dictated by the fact that we were limited to 400-mW pump power in our laboratory, which impacts the range of nonlinear phase shifts that can be reached conveniently in a controlled experiment. The pulse is stretched in 100 m of fiber. After a preamplifier stage, the repetition rate can be cut from 40 MHz to 3 MHz with an acousto-optic modulator (AOM). The amplified and compressed pulses are characterized with an interferometric autocorrelator. Fig. 1 shows that the autocorrelation obscures the dramatic variation in the pulse shape and asymmetry as $\Phi^{NL}$ is varied. However, the variation in the pulse duration is readily observable. More-detailed information will be obtained by recording the cross-correlation of the compressed pulse with the input pulse, or by recording the intensity and phase with frequency-resolved optical gating [14], e.g., and these will be the subject of future work.

The variation of the output pulse with $\Phi^{NL}$ is shown in Fig. 4. The lowest pulse energy was limited by the desire to record the autocorrelation with adequate signal-to-noise ratio. With the AOM turned off, the pulse energy was 3 nJ, which produced $\Phi^{NL} = 0.4\pi$. (The $\Phi^{NL}$ obtained with a given pulse energy is slightly larger than that of the simulations, which neglect ~2 m of SMF that couple light into and out of the AOM.) The autocorrelation (Fig. 4(a)) implies a pulse duration of 240 fs. A similar result is obtained with lower signal-to-noise ratio when the AOM is turned on to reduce the repetition rate. At 3 MHz, the pulse is amplified to 15 nJ, and $\Phi^{NL} = 1.8\pi$. The pulse duration (Fig. 4(b)) decreases to 180 fs, and the secondary structure that arises from TOD diminishes. When the pulse energy increases to 17 nJ ($\Phi^{NL}=2.1\pi$), the pulse duration increases again to 195 fs and the secondary structure begins to increase as well. The experimental trend agrees qualitatively and semi-quantitatively with the numerical simulations, and exhibits a clear minimum in the pulse duration near the expected optimum value of $\Phi^{NL}$. As expected, the power spectrum broadens slightly at the highest pulse energies.

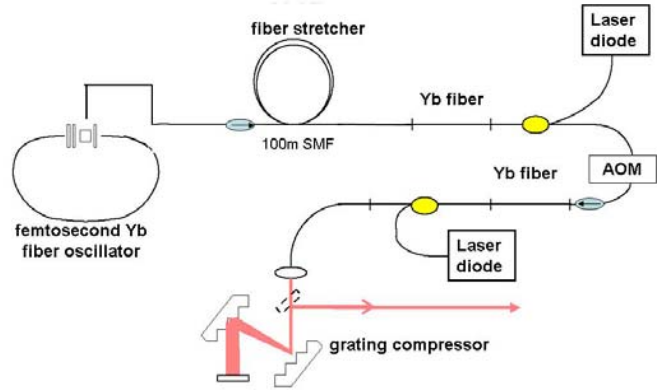

Figure 3. Schematic of the experimental setup.

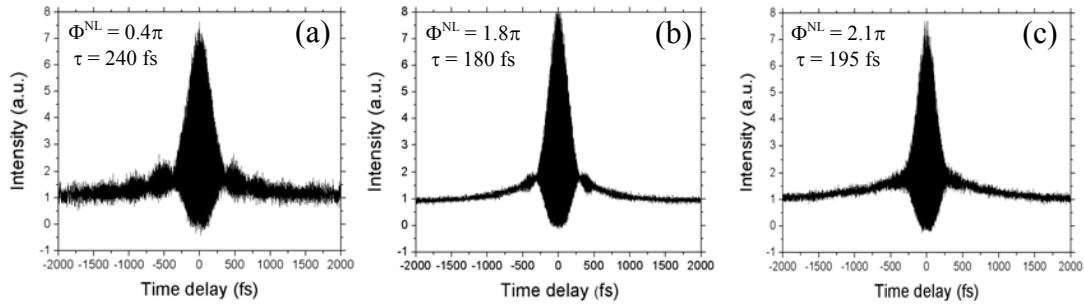

Figure 4. Autocorrelations of amplified and dechirped pulses with a) $\Phi^{NL} = 0.4\pi$, b) $\Phi^{NL} = 1.8\pi$, and c) $\Phi^{NL} = 2.1\pi$. The corresponding pulse duration is shown in each panel.

### 4. Discussion and conclusion

The demonstration that TOD and nonlinear phase shifts can compensate each other seems likely to change the way future short-pulse fiber amplifiers are designed. The agreement between the experimental results and the numerical simulations allows us to extrapolate to higher pulse energies than employed here. There is great interest in construction of fiber amplifiers for microjoule and even millijoule-energy pulses, but nonlinearity presents a major challenge in the design of such devices. *In the approach presented here, nonlinearity is desirable.* Amplifiers can be designed to operate with a certain nonlinear phase shift, instead of attempting to avoid nonlinearity entirely. The stretcher and compressor can be designed to operate with a certain TOD, instead of attempting to minimize it. Our initial calculations indicate that microjoule-energy pulses of ~300-fs duration can be obtained from amplifiers that use only SMF, and millijoule-level pulses may be obtained with multimode fiber [11] or large-mode-area photonic-crystal fiber [10]. Scaling to shorter pulses is also possible. The overwhelming practical benefits of fiber stretchers would recommend their use if the resulting performance was equal to, or even slightly worse than, that obtained with a grating stretcher. With the possibility of better performance, fiber stretchers should become the standard in short-pulse fiber amplifiers. The compensation of nonlinearity by TOD can be implemented in various ways. Fiber Bragg gratings offer an appealing combination of compactness and dispersion control and efforts exist to demonstrate customized chirping of the grating period to compensate higher orders of dispersion. Disadvantages of FBGs include an inherent tradeoff between bandwidth and dispersion, the requirement of expensive (often custom) fabrications, and limited adjustability in dispersion for a given design. These properties contrast with the simplicity and availability of SMF. Of course, the compensation of nonlinearity by TOD described here can be implemented with FBGs: the FBG would be designed to provide a certain amount of TOD, depending on the desired pulse energy. Similar arguments can be made for dechirping the pulse with photonic-bandgap fiber.

To summarize, we have demonstrated that nonlinearity and TOD can compensate each other to a large degree in CPA systems. For a given magnitude of TOD, there exists an optimum value of the nonlinear phase shift, for which the output pulse duration is minimized. The output pulse can be significantly shorter and cleaner than in the absence of nonlinearity, and the peak power is correspondingly increased. Initial experiments with low-energy (~20 nJ) femtosecond pulses clearly exhibit the main features and trends of the theoretical predictions. Extension of this approach to higher pulse energies appears to be straightforward. Significantly, the benefits of this approach increase with nonlinear phase shift, *i.e.*, with pulse energy. Compensation of dispersion beyond third-order by nonlinear phase shifts may also be expected. Finally, the concept described here can be combined with other devices such as fiber Bragg gratings or photonic-bandgap fibers. We expect this technique to find significant use in future high-energy short-pulse fiber amplifiers.

**Acknowledgements**

Portions of this work were supported by the National Science Foundation (ECS-0217958) and DARPA. The authors acknowledge valuable discussions with J. Moses, Y.-F. Chen, F. Ilday, H. Lim, and T. Sosnowski.